\documentclass[aps,prb,preprint,twocolumn,10pt,floatfix]{revtex4-2}
\usepackage{CJK}
\usepackage{float}
\usepackage{graphicx,amsmath,amssymb,bm}
\usepackage{color}
\usepackage{appendix}
\usepackage{ulem}

\begin{document}
\begin{CJK*}{GB}{}

\title{Imaginary phonon modes and phonon-mediated superconductivity in Y$_2$C$_3$}

\author{Niraj K. Nepal$^1$}
\author{Paul C. Canfield$^{1,2}$}
\author{Lin-Lin Wang$^1$}
\email[]{llw@ameslab.gov}
\affiliation{[1] Ames National Laboratory, Ames, Iowa 50011, USA}
\affiliation{[2] Department of Physics and Astronomy, Iowa State University, Ames, Iowa 50011, USA}

\date{\today}

\begin{abstract}

For Y$_2$C$_3$ with a superconducting critical temperature (T$_c$) $\sim$18 K, zone-center imaginary optical phonon modes have been found for the high-symmetry $I$-$43d$ structure due to C dimer wobbling motion and electronic instability from a flat band near Fermi energy. After lattice distortion to the more stable lowest symmetry $P1$ structure, these stabilized low-energy phonon modes with a mixed C and Y character carry a strong electron-phonon coupling to give arise to the observed sizable T$_c$. Our work shows that compounds with the calculated dynamical instability should not be simply excluded in high-throughput search for new phonon-mediated superconductors. Moreover, we have studied the phase stability of the $I$-$43d$ structure by calculating the enthalpy of different structural motifs of binary compounds containing group IV elements at the 2:3 composition and also exploring the energy landscapes via $ab$ $initio$ molecular dynamics near and out of the $I$-$43d$ structure. Our results show that the $I$-$43d$ type structures with C dimers are preferred in the low to medium pressure range. Because of the wobbling motion of the C dimers, there are many local energy minimums with degenerated energies. Thus, the ensemble average of many $I$-$43d$-distorted structures with C dimer wobbling motion at finite temperature still gives an overall $I$-$43d$ structure.
\end{abstract}
\maketitle

\end{CJK*}

\section{Introduction}
Computational search for new phonon-mediated superconducting (SC) compounds has become an active field since the discovery of SC critical temperature (T$_c$) near room temperature for metal hydrides under high pressure ($\sim$200 GPa) \cite{LHLLM14,WMH15, DETKS15,P23,LHHPSZPADE22,CG22}. The search requires the computation of electron-phonon coupling (EPC) matrices within the density functional perturbation theory (DFPT) \cite{BDDG01,D01} to estimate T$_c$ using Migdal-Eliashberg approximations \cite{M58,E60,A72,AD72,MG13}. However, compounds with calculated imaginary phonon modes are regarded as dynamically unstable and often discarded as promising candidates in high-throughput\cite{CG22}. As the phonon eigenmodes are solved in the quasi-harmonic approximation, if some diagonal elements of the dynamical matrix are negative, the square root will give imaginary values for frequency, which correspond to local maximum instead of minimum for the potential energy surface. It means the total energy of the system can be lowered via distortion by following the eigenvectors of these imaginary phonon modes, for which the frequencies are usually plotted as negative values. On the other hand, some well-known SC are metastable, such as YPd$_2$B$_2$C \cite{CTEZ94,SRMCGP94}. In this study, using yttrium sesquicarbide (Y$_2$C$_3$) as an example among rare-earth carbides R$_2$C$_3$ (R = rare-earth, Y and La) with their SC being discovered and improved over the years \cite{KGKS69,AAMZA04,SSSTO07,KSAHMSTK08,CSAASZJY11,AOKMGMA08}, we find that the high-symmetry body-centered cubic (BCC) Pu$_2$C$_3$-type structure, Pearson symbol cI40 in space group $I$-$43d$ (220), is dynamically unstable with zone-center imaginary optical phonon modes, which once stabilized carry a large EPC strength ($\lambda$) to give the sizable T$_c$. Thus, overlooking new compounds with the similar imaginary phonon modes in high-throughput search can leave out promising SC candidates.

Y$_2$C$_3$ in the space group $I$-$43d$ (220) with C dimers was first synthesized in 1969 using high-pressure, high-temperature techniques\cite{KGKS69} to show T$_c$ ranging from a low of 6 K to a high of 11.5 K \cite{KGKS69}. Subsequent experiments with Thorium (Th) alloying increased the T$_c$ to 17 K in (Y$_{0.7}$Th$_{0.3}$)$_2$C$_3$ \cite{KGKS69_th}. Recently in 2004, for the samples prepared under 4.0-5.5 GPa followed by different heat treatment and sintering conditions, a T$_c$ up to 18 K in binary Y$_2$C$_3$ has been observed \cite{AAMZA04}. Electronic structure calculations have found that the states near the Fermi energy (E$_F$) are hybridized \cite{NSO07,SI04,SM04} between Y 4$d$ and C-C antibonding 2$p$ orbitals with an interesting flat band \cite{NSO07,SM04} close to the E$_F$, as well as highly degenerated Kramers-Weyl points at $\sim$1 eV below the E$_F$ \cite{JLZDL21}. It has been proposed that the sensitivity of SC properties for the related La$_2$C$_3$ \cite{KWRVO07} with respect to synthesis condition is due to the change of density of states (DOS) at the E$_F$ from C deficiency\cite{SI04,SM04}. An earlier EPC calculation\cite{SM04,P08} on Y$_2$C$_3$ has focused only on two zone-center modes; a Y-dominated mode at 175 cm$^{-1}$ (5.2 THz) and the C-C bond stretching mode at 1442 cm$^{-1}$ (43.3 THz), and found a much larger EPC contribution from the Y-dominated than C-C bond stretching mode. However, here with DFPT calculation over all the phonon modes, we find that the high-symmetry $I$-$43d$ structure is dynamically unstable with zone-center imaginary phonon modes of C dimer wobbling motion due to electronic instability from the flat band along the $\Gamma$-N direction at E$_{F}$. By following the lattice distortion of the imaginary modes with full relaxation, more stable low-symmetry structures of Y$_2$C$_3$ have been found, where the imaginary modes become stabilized in a low-energy range, lower than the Y-dominated modes. These stabilized low-energy phonon modes carry large $\lambda$, which contributes significantly to T$_c$. The imaginary phonon modes can also be stabilized with a large electronic smearing and under pressure, with the latter enhancing the T$_c$ in a trend agreeing with the previous experiment \cite{AAMZA04}.

\section{Computational Details}

 We used quantum espresso (QE) package \cite{QE09,QE217} with ultrasoft pseudopotentials \cite{SSSP18, GBRV14} for ground-state and EPC calculations. The exchange-correlation energy was approximated by Perdew-Burke-Ehrnzerhof (PBE) functional \cite{PBE96}. Kinetic energy cutoff of 50 Ry, Brillouin zone (BZ) sampling of (6$\times$6$\times$6) \textbf{k}-point mesh, and a Gaussian smearing of 0.05 eV were utilized for ground-state calculations. Phonon calculations were performed with momentum transfer grid (\textbf{q}) of (2$\times$2$\times$2). Further \textbf{k}-mesh and \textbf{q}-mesh were interpolated to (12$\times$12$\times$12) fine-mesh to compute EPC properties. The Coulomb potential $\mu_c^*$ was kept fixed at 0.16. The Fermi surfaces were plotted using the Fermi surfer package \cite{fermisurfer}.  The Birch-Murnaghan equations were employed for the calculation of the equation of state (EOS) \cite{birch47,Murn44}. The formation energy per atom was computed as,
 \begin{equation}
     \Delta E_f = E_{Y_2C_3} - \frac{2}{5}E_Y -  \frac{3}{5}E_C,
 \end{equation}
 where E$_i$'s are the total energy per atom of the compound Y$_2$C$_3$ and its constituents, namely, Y in hexagonal closed packed and C in the diamond structure. For $ab$ $initio$ molecular dynamics (MD) and phase stability calculations with different structural motifs, we used Vienna $ab$ $initio$ simulation package (VASP) \cite{KF96,KF96b} and projector augmented wave (PAW) method \cite{B94} with a kinetic energy cutoff of 500 eV. The MD simulations with $NpT$ ensembles were first performed on the $I$-$43d$ primitive cell of Y$_2$C$_3$ to search for new structures at different pressure near and out of the energy minimum of $I$-$43d$, and then used on the (2$\times$2$\times$2) supercell of the conventional cell with 320 atoms to explore the degeneracy of the C dimers orientation due to wobbling motion. From the MD trajectories, the snapshots of the low potential energy configurations were selected for full relaxation \cite{WJD07}.

\section{Results and Discussions}
\begin{figure}[h!]
    \centering
    \includegraphics[scale=0.28]{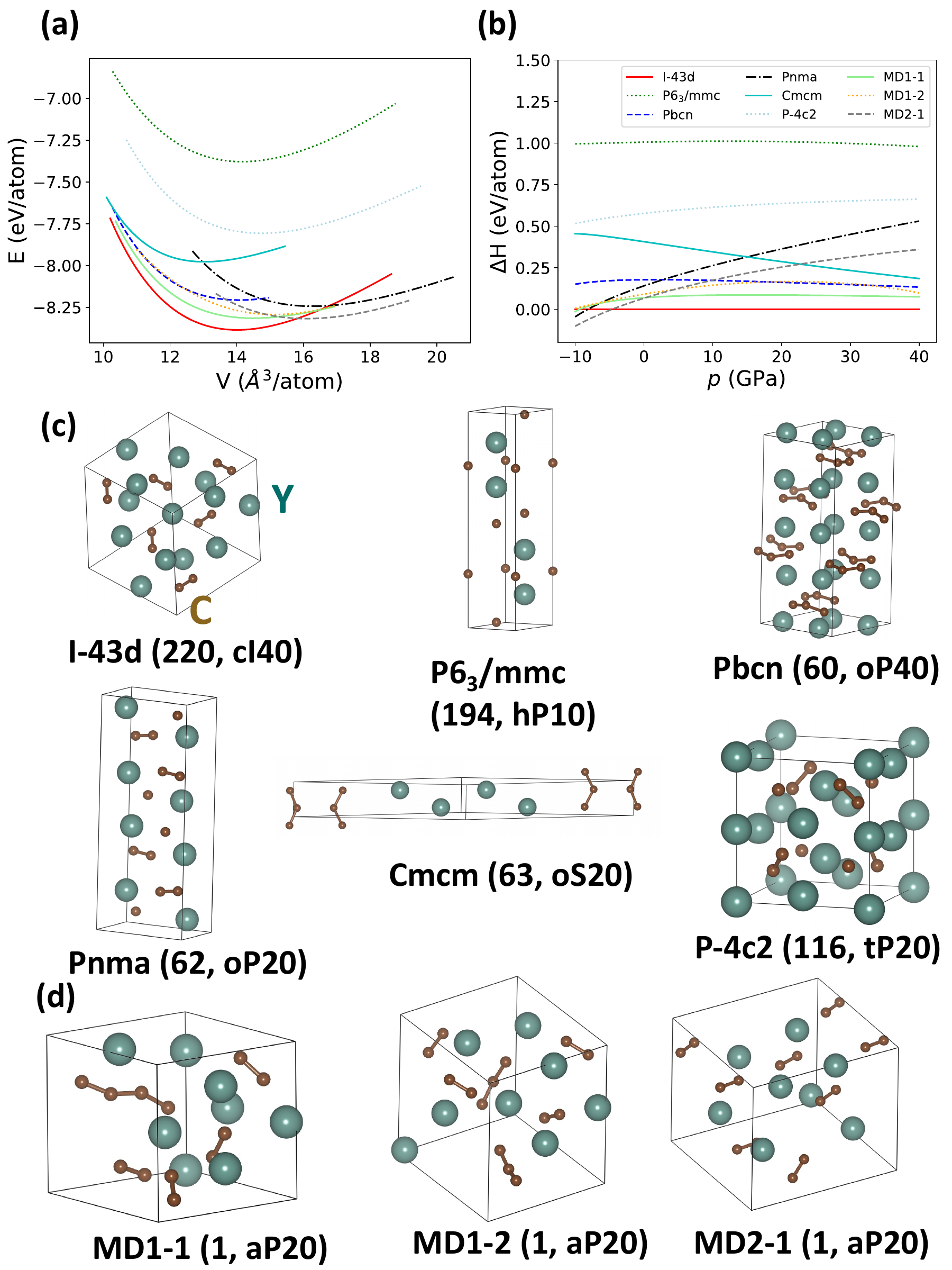}
    \caption{(a) Energy ($E$) vs volume ($V$) plots for Y$_2$C$_3$ in the various structural motifs of the binary compounds containing group IV elements at the composition of 2:3 and the structures close to $I$-$43d$ phase obtained from $ab$ $initio$ molecular dynamics (MD) simulations. (b) Enthalpy difference ($\Delta H$) calculated for each phase vs pressure ($p$) with respect to the $I$-$43d$ phase. (c) Fully relaxed Y$_2$C$_3$ in the different structural motifs as plotted in (a) and (b). (d) Fully relaxed Y$_2$C$_3$ structures obtained from the MD simulations under pressure of 20 GPa (MD1-1 and MD1-2) and $-$5 GPa (MD2-1). The space group number and Pearson symbol are also shown in parenthesis.}
    \label{fig:0}
\end{figure}
\begin{figure}[h!]
    \centering
    \includegraphics[scale=0.35]{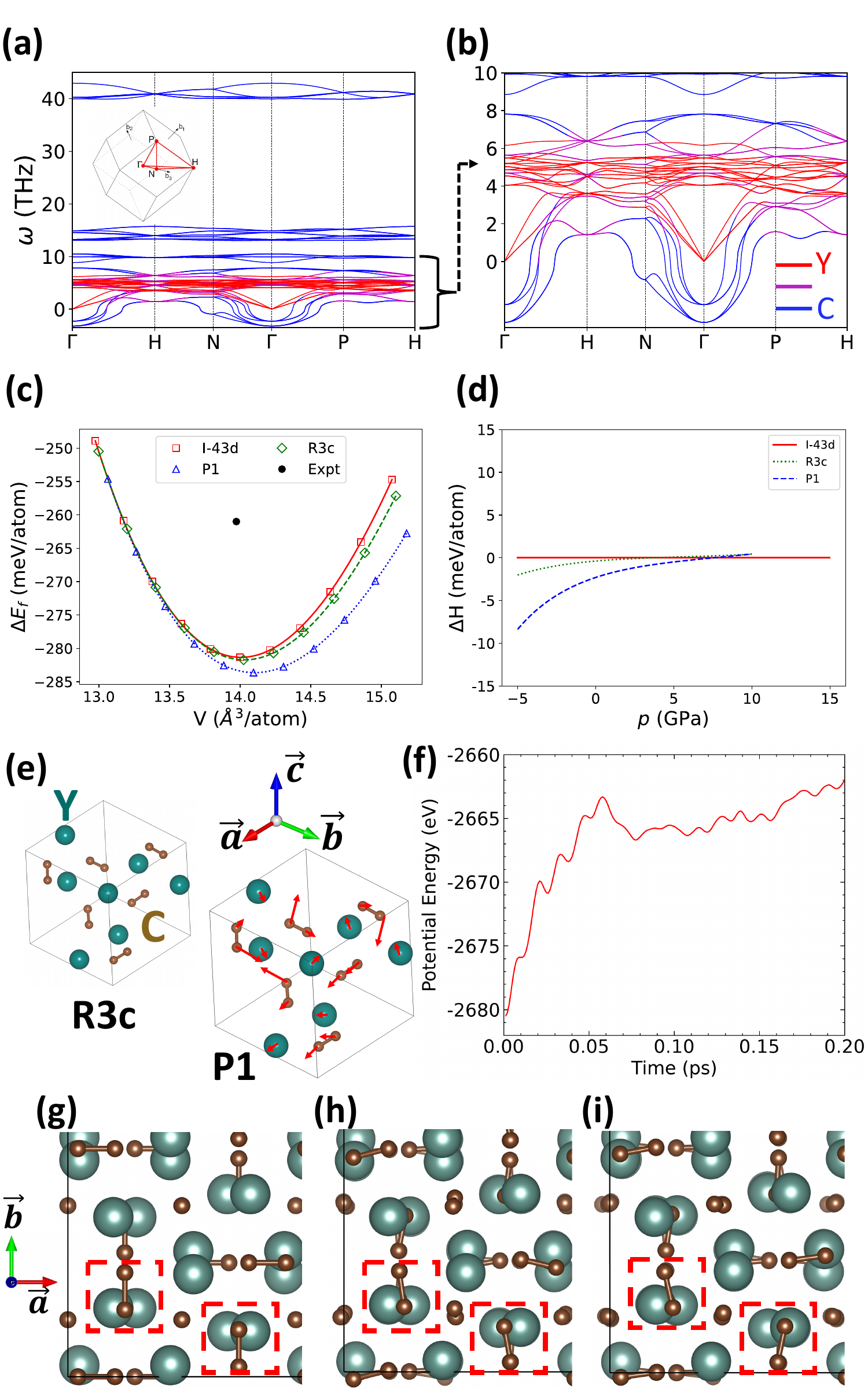}
    \caption{(a) Atom-projected phonon dispersion of $I$-$43d$ structure exhibiting imaginary modes shown as negative frequencies ($\omega$). First Brillouin zone (BZ) and high-symmetry paths are in inset. (b) Atom-projected phonon dispersion $<$10 THz zoomed from (a). Red represents for Y, blue for C, and purple for contributions from both as defined in Eq.~\ref{eq1}. (c) Formation energy ($\Delta$E$_f$) with respect to volume ($V$) for $I$-$43d$ (220), rhombohedral structure in space group $R3c$ (161), and the low-symmetry structure relaxed from eigenmode 4 in space group $P1$ (1), compared to the experimental data {\color{blue}\cite{N80}} around the equilibrium. (d) Enthalpy difference ($\Delta H$ vs $p$ plot computed for the structures with respect to $I$-$43d$ structure. (e) Primitive unit cell of Y$_2$C$_3$ in the $R3c$ and $P1$ structures. For $P1$, the structure is presented along the [111] direction with atomic displacements represented by {\color{blue}red} arrows according to the imaginary eigenmode 4. (f) Potential energy vs time plot for \textit{ab initio} MD simulations. (g) Zoomed-in sections of the supercell of 320 atoms in the $I$-$43d$ structure and (h)-(i) those distorted structures having degenerate energy with different C dimer orientations obtained from the MD simulations. The different C dimer orientations due to wobbling motion are highlighted in the dashed red squares.}
    \label{fig:1}
\end{figure}
In Figure.~\ref{fig:0}, we first establish that the BCC-type $I$-$43d$ structure is the ground-state phase in the moderate pressure range. (We will use BCC and $I$-$43d$ interchangeably.) In Fig.~\ref{fig:0}(a), we have plotted the total energy ($E$) with respect to volume ($V$) per atom for different phases. First, we have considered all the binary compound structural types containing group IV elements at the 2:3 composition (Fig.1(c)) and found several structural motifs of C atoms ($P6_3/mmc$), C trimers ($Pbcn$), zig-zag C chains ($Cmcm$), and mixed C dimers and atoms ($Pnma$ and $P$-$4c2$). Then, we also used $ab$ $initio$ MD to explore energy landscape  \cite{WJD07} near and out of the $I$-$43d$ structure at 20 GPa (MD1) and $-$5 GPa (MD2) for possible other low-symmetry structures (Fig.~\ref{fig:0}(d)). We found structures of mixed C dimer and tetramer (MD1-1), mixed C dimer and trimer (MD1-2) and C dimers in a different lattice (MD2-1). As shown by the $E(V)$ EOS in Fig.~\ref{fig:0}(a), for a large range of volume from 11 to 16 \AA$^3$/atom, the $I$-$43d$ structure has the lowest energy, and only for volume larger than the equilibrium of the $I$-$43d$ (negative pressure), two structures ($Pnma$ and MD2-1) become more stable. With the $E(V)$ data, we computed the enthalpies ($H=E+pV$) for different pressure ($p$) relative to the $I$-$43d$ phase and present in Fig.~\ref{fig:0} (b). Except for negative pressure less than $-$5 GPa (expanding volume), the $I$-$43d$ phase is preferred up to 40 GPa, which is the range we are focusing in the current study to stabilize the imaginary phonon modes to reveal their large EPC contribution for enhancing T$_c$ in comparison to the relevant experiments. From the slope of $H(p)$, there are possible structural phase transitions at pressure well above 100 GPa, but it is beyond the scope of the current paper and needs to be further explored in a future work.

\begin{figure}[h!]
    \centering
    \includegraphics[scale=0.7]{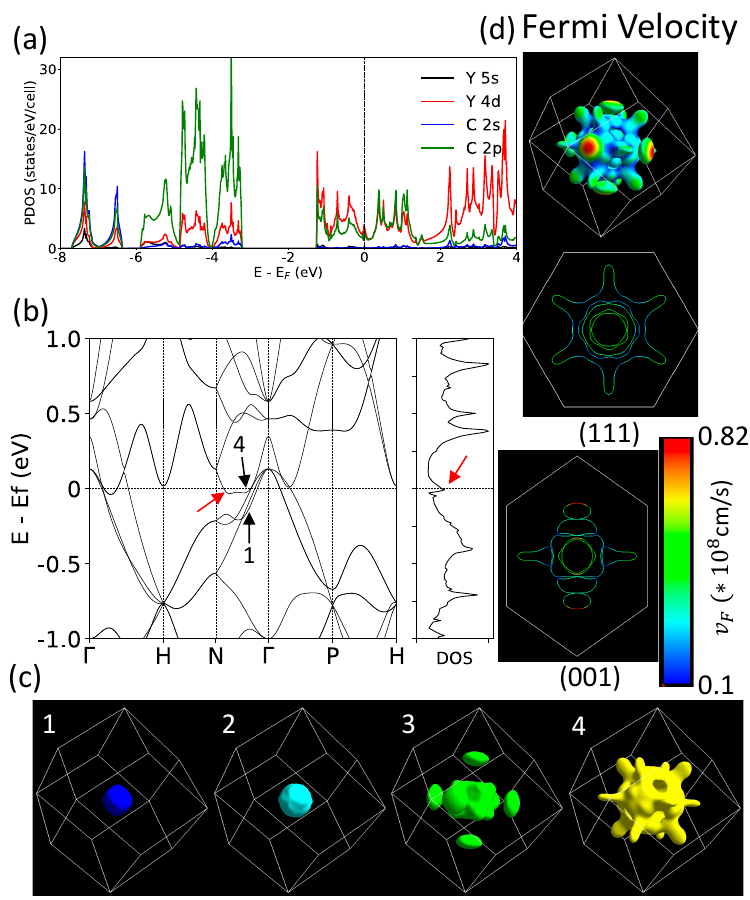}
    \caption{(a) Projected density of states (PDOS) of Y$_2$C$_3$ in the $I$-$43d$ structure on Y 5$s$, Y 4$d$, C 2$s$ and C 2$p$ orbitals, showing a DOS peak near the Fermi-level (E$_{F}$). (b) Electronic band structure in the range of E$_{F}\pm$1 eV showing four bands crossing the E$_{F}$ (3 valence and 1 conduction bands indicated by black arrow, indexed from ``1" to ``4" at the top). The flat band section is along the $\Gamma$-N direction (4$^{th}$ band indicated by red arrow). (c) Fermi surface (FS) corresponding to the four different bands, labelled by their band index defined in (b). (d) Fermi velocity (\textbf{v}$_F$) plotted on the FS. 2D FS contour on the (111) and (001) planes are also shown.}
    \label{fig:2}
\end{figure}

Figure~\ref{fig:0}(c) shows the $I$-$43d$ structure of Y$_2$C$_3$ in the primitive unit cell (20 atoms), where eight Y atoms occupy 16$c$ sites and twelve C atoms at 24$d$ sites form six dimers. Our PBE-relaxed lattice constant of 8.239\AA\ and C-C bond length of 1.34\AA\ agree well with the experimental 8.237\AA\ and 1.298\AA\ \cite{MNKTHI05}, and also 8.254\AA\ and 1.33\AA\ from the previous DFT calculation\cite{NSO07}, respectively. The DFPT-calculated $I$-$43d$ phonon dispersion is plotted in Fig.~\ref{fig:1}(a) with the inset showing the Brillouin zone (BZ) and high-symmetry points. To analyze the contribution of different atomic species to phonon eigenmodes, we have summed the atomic displacements of each eigenmode for different species ($|\textbf{e}^i_{\nu \textbf{k}}|^2$) as in Eq.(~\ref{eq1}) and color-coded the dispersion,
\begin{equation}
    |\textbf{e}^i_{\nu \textbf{k}}|^2 = \sum_{j=1}^{N_i}  |\textbf{e}^{ij}_{\nu \textbf{k}}|^2,
    \label{eq1}
\end{equation}
where $\textbf{e}^{ij}_{\nu \textbf{k}}$ are the displacements obtained for atom type $i$ from the eigenvector of mode $\nu$ for $k$-point $\textbf{k}$ by diagonalizing dynamical matrix, and index $j$ runs to $N_i$ as the number of atoms for type $i$. For Y$_2$C$_3$ in Fig.~\ref{fig:1}(a), the modes with the highest frequency $\sim$40 THz are well separated from the rest and solely contributed from the C dimer stretching mode. Our atomic projection also reveals that the modes from 10 to 15 THz are dominated by other C-related motion (blue). In contrast, the Y contributions (red) are mostly around 5 THz due to the much larger mass of Y than C. As zoomed in Fig.2(b) for the frequency $<$10 THz, there are modes with mixed contributions from both Y and C (purple). Interestingly, we find three zone-center modes at $\Gamma$ are imaginary shown by the negative frequency and are also dominated by C contributions. Figure~\ref{fig:1}(e) shows one of imaginary modes (eigenmode 4) with the arrows representing the direction and size of the atomic displacements for the eigenvector. Usually the C-related modes are assumed to have high frequency due to the tendency of C to form stronger covalent bonds and smaller mass than Y. But as shown in the Appendix Fig.A1(a)-(c) with the conventional unit cell, each Y is surrounded by 12 C dimers in a twisted prism, while each C dimer is surrounded by eight Y atoms in a cage of two trapezoids. The wobbling motion of the C dimers within these cages have zero to even negative energy cost, giving arise to the three imaginary optical modes. The other two imaginary modes (eigenmode 1 and 10) are displayed in Appendix Fig.A2.

The zone-center imaginary phonon modes indicate that the $I$-$43d$ structure of Y$_2$C$_3$ is dynamically unstable. To search for more stable structures, we have used the imaginary modes to distort the primitive unit cell to perform full relaxation with atomic positions and cell vectors, and we find new low-symmetry structures with total energies of 2.34 (mode 4 and 10) and 0.9 meV/atom (mode 1) below that of the $I$-$43d$ structure. For the most stable structure of mode 4, the lattice symmetry is also the lowest in space group $P1$ (1). We have also considered the cubic to rhombohedral distortion along the [111] direction to retain some symmetry as in space group $R3c$ (161). In Fig.~\ref{fig:1}(c), the formation energy ($\Delta$E$_f$) is plotted with respect to $V$ close to the equilibrium for the three structures with gradually lowered symmetry. The $\Delta$E$_f$ of $-281$meV/atom agrees well with the experimental $-261$meV/atom \cite{N80}. The relaxed C-C (Y-C) bond length in the $I$-$43d$ structure around 1.34 (2.51)\AA\ is consistent with previous DFT calculation \cite{NSO07}, and also comparable to the experimental 1.298 (2.51)\AA\ \cite{MNKTHI05}. Compared to the high-symmetry $I$-$43d$ structure, the $R3c$ and $P1$ structures are more stable at the equilibrium and larger volumes with $P1$ being the most stable in the lowest symmetry. But at volumes smaller than the equilibrium or under increasing pressure, the three EOS $E(V)$ plots converge to the same curve. In other words, $I$-$43d$ structure becomes the ground state under pressure.

In addition to the $E(V)$, We also plot $\Delta H(p)$ in Fig.~\ref{fig:1}(d), which shows the low-symmetry BCC-distorted structures have lower $H$ than the high-symmetry $I$-$43d$ only in the small pressure range below 7 GPa. However, we would like to point out a major difference between these BCC-distorted structures and the various structures presented in Fig.~\ref{fig:0}, because at positive pressure and smaller volume, the EOS of the BCC-distorted structures merge into that of BCC. The underlying reason is that these distortions are from the wobbling motion of C dimers in the Y cages of $I$-$43d$, which can have many similarly distorted structures with degenerated energy. Using \textit{ab initio} MD simulation of a supercell of 320 atoms to explore the trajectories of C dimer wobbling motion (Fig.~\ref{fig:1}(f)), we can then take snapshots of the configurations for full ionic relaxation. For example, we show two configurations with degenerated energy have C dimers in different orientations in Fig.~\ref{fig:1}(h) and (i). Thus, the ensemble of the many BCC-distorted structures with C dimer wobbling motion at finite temperature should still give an overall BCC structure as seen in X-ray diffraction. This type of metastable BCC Y$_2$C$_3$ structure is similar to the austenitic NiTi phase, for which the high-symmetry CsCl (B2) structure has imaginary phonon modes and multiple local energy minimums, thus, the overall stable B2 structure is represented by distorted supercells of the B2 structure \cite{ZJ14}.

\begin{figure}[h!]
    \centering
    \includegraphics[scale=0.25]{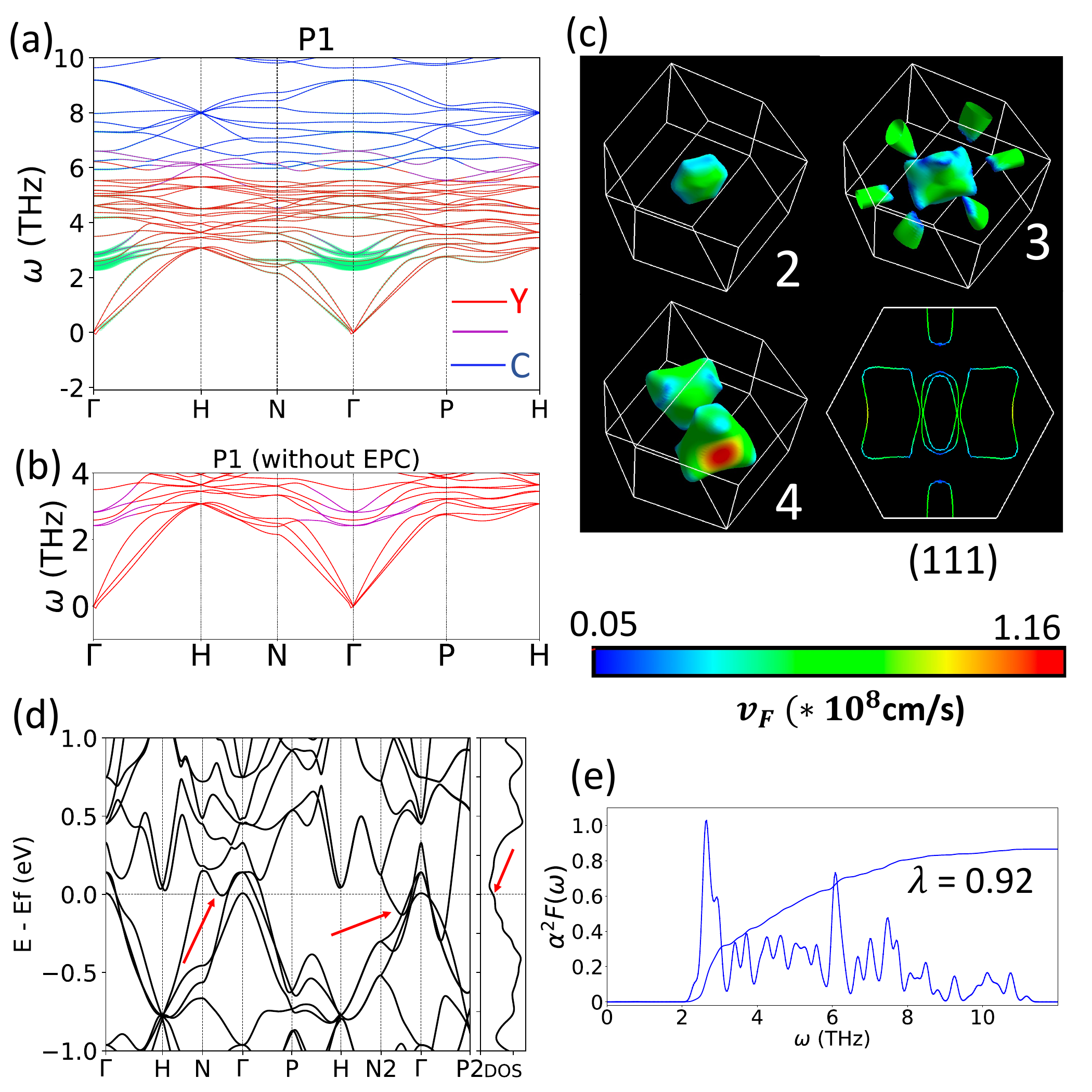}
    \caption{(a) Atom-projected phonon dispersion for low-symmetry $P1$ structure of Y$_2$C$_3$ relaxed from eigenmode 4. Atom-projections are highlighted by red for Y, blue for C, and purple for both contributions. The mode-resolved electron-phonon coupling (EPC) strength ($\lambda$) is represented by green shade. (b) Atom-projected phonon dispersion without the shade of EPC zoomed $<$4 THz to show the contributions from both Y and C atoms to the $\lambda$. (c) Fermi velocity (\textbf{v}$_F$) plotted over the Fermi surfaces (FS), which are indexed with the same numbers as in Figure.~\ref{fig:2} but become distorted because of the relaxation. The 2D FS contour on (111) plane is also shown. (d) Electronic band structure of $P1$ structure showing the absence of flat band, thereby flattening the density of states peak (red arrows) at the Fermi-level (E$_{F}$). (e) Isotropic Eliashberg spectral function ($\alpha^2F(\omega)$) of $P1$ structure gives the overall integrated $\lambda$=0.92.} 
    \label{fig:3}
\end{figure}

To find out the origin of such instability in BCC Y$_2$C$_3$, we have calculated the electronic structures. As plotted in Fig.~\ref{fig:2}(a), the partial DOS (PDOS) projected on atomic orbitals shows that the bonding states of C dimer with 2$s$ and 2$p$ orbitals dominate from $-8$ to $-3$ eV, while the Y 4$d$ orbitals dominate the empty states above +2 eV. In the middle, the hybridization between the C dimer antibonding states and Y 4$d$ form a pseudo-gap from $-0.5$ to +0.5 eV. However, inside this broad pseudo-gap, there is a sharp local DOS maximum at E$_F$, which hints at an electronic instability. From the calculated electronic band structures in Fig.~\ref{fig:2}(b) in the range of $\pm$1 eV, there are four bands crossing E$_F$ with the corresponding 3D Fermi surface (FS) plotted in panel (c). As numbered from low to high energy in Fig.~\ref{fig:2}(b-c), the first two valence bands form small spherical hole pockets around $\Gamma$. The third and highest valence band has disc-like features along the $\Gamma$-H direction besides a cubic hole pocket. The 4th and lowest conduction band forms spike-like features along the $\Gamma$-N direction besides a spherical double-layered shell around $\Gamma$. The spike-like features come from the flat band along the $\Gamma$-N direction in (b) and corresponds to the sharp local DOS maximum at $-10$ meV (red arrow). The Fermi velocity (\textbf{v}$_F$), $\nabla_k E_k$, has also been calculated and overlaid on FS in panel (d). The maximum value appears on the outer surface of the disc-like feature of FS3. 2D cuts of FS along (111) and (001) planes show that the distribution of \textbf{v}$_F$ with high values across FS1 and FS2, relatively lower value for FS3 except for the disc-like feature, and the lowest for FS4 (blue and green patches), consistent with previous result\cite{NSO07}. Although the presence of the flat band in Y$_2$C$_3$ has been noted before (Refs.~\cite{NSO07} and \cite{P08}), its implication on the structural instability has only been revealed in our current work by finding the zone-center imaginary phonon modes and the more stable structures via distortion according to the imaginary phonon modes. The flat band in Y$_2$C$_3$ is present in the high-symmetry BCC structure not the low-symmetry BCC-distorted structure at ambient pressure. Our Fig.A3(a) shows the flat band crosses the E$_F$ with increasing pressure in the pressure-stabilized BCC structure. Thus, experimental observation of the flat band in Y$_2$C$_3$ requires angular-resolved photo-emission spectroscopy (ARPES) experiment on single crystal samples of Y$_2$C$_3$ at elevated pressure.

After establishing the electronic origin of the instability in BCC Y$_2$C$_3$ and finding the more stable low-symmetry $P1$ structure, next we calculate EPC in DFPT to explain SC properties. In Figure.~\ref{fig:3}, we plot the phonon dispersion of the $P1$ structure relaxed from eigenmode 4 ($\sim$2.34 meV/atom more stable than $I$-$43d$) in (a) with and (b) without the mode-resolved EPC projections. Atomic characters are also similarly projected as in Fig.~\ref{fig:1}(d-e). Due to the lowest symmetry, many degeneracy at the high-symmetry points have been lifted in $P1$ (Fig.~\ref{fig:3}(a)) when compared to the $I$-$43d$ phonon dispersion (Fig.~\ref{fig:1}(e)). Importantly, the $P1$ structure does not have imaginary modes and is dynamically stable. The original imaginary modes in $I$-$43d$ are now stabilized in the low frequency range of 2-3 THz at $\Gamma$. Moreover, the largest EPC contribution is found around $\Gamma$ (green shade) along the $\Gamma$-H, $\Gamma$-N, and $\Gamma$-P directions within the same low frequency range, where both C dimers and Y atoms provide significant contributions (see purple in Fig.~\ref{fig:3}(b) without EPC). As confirmed in the Eliashberg spectral function of Fig~\ref{fig:3}(e), the largest EPC contribution peak comes from the 2-3 THz range, followed by another peak around the 6 THz, both have the mixed C and Y characters, giving an overall EPC strength $\lambda$ of 0.92, logarithmic average of phonon frequencies ($\omega_{log}$) of 219.6 K, and T$_c$ of 9.3 K as listed in Table~\ref{tab1}. The calculated T$_c$ is in a good agreement to the experimental data. For the change in electronic structures, the FS of the $P1$ structure now consists of only three bands, completely removing the spike-like features due to the flat band along the $\Gamma$-N direction [Fig.~\ref{fig:3}(b)]. As the result, the electronic instability associated with the flat band is eliminated, and the local DOS maximum at E$_F$ is flattened out, as seen in Fig.~\ref{fig:3}(c) (red arrows). The results from the other $I$-$43d$ imaginary phonon modes, namely mode 1 ($\sim$ 0.9 meV/atom lower) and 10 ($\sim$ 2.34 meV/atom lower, similar to eigenmode 4) are presented in Fig.A2. The corresponding properties are listed in Table~\ref{tab1} showing a similar T$_c$ to the $P1$ structure of eigenmode 4. Thus, one of the key findings of our work is that the EPC for the phonon-mediated SC in Y$_2$C$_3$ with a sizable T$_c$ arise from the stabilized imaginary phonon modes after removing the flat band and electronic instability, and the largest EPC contributions are from these stabilized low-energy optical phonon modes of mixed C and Y characters.

\begin{figure}[h!]
    \centering
    \includegraphics[scale=0.3]{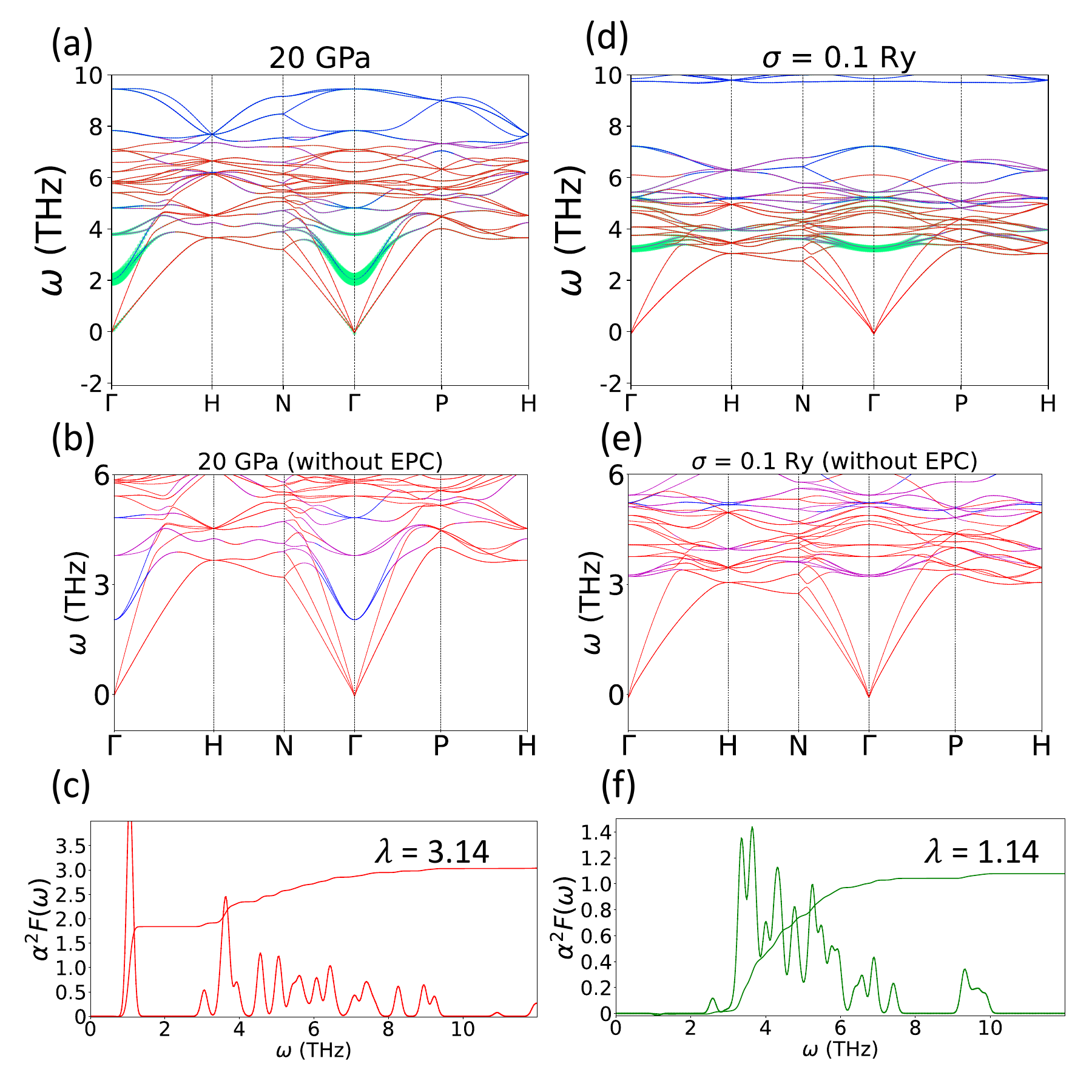}
    \caption{(a) Atom-projected phonon dispersion of Y$_2$C$_3$ in the $I$-$43d$ structure under 20 GPa. Atom-projections are highlighted by red for Y, blue for C, and purple for both contributions. The mode-resolved electron-phonon coupling (EPC) strength ($\lambda$) is represented by green shade. (b) Atom-projected phonon dispersion without the shade of EPC zoomed around the modes with the largest EPC to show the contributions from both Y and C atoms to the $\lambda$. (c) Isotropic Eliashberg spectral function ($\alpha^2F(\omega)$) of the $I$-$43d$ structure under 20 GPa giving the overall $\lambda$=3.14. (d)-(f) Similar plots to (a)-(c) with a large electronic smearing of 0.1 Ry at zero pressure. The overall integrated $\lambda$=1.14. The spectral function ($\alpha^2F(\omega)$) is computed in isotropic Eliashberg approximation and then integrated (dashed lines) for the overall $\lambda$.}
    \label{fig:4}
\end{figure}

\begin{table}[h!]
    \centering
     \caption{EPC strength $\lambda$, logarithmic average of phonon frequencies $\omega_{log}$, and critical temperature T$_c$ computed from isotropic Eliashberg approximation with distorted structures, pressure, and large electronic smearing for Y$_2$C$_3$ using the Coulomb potential $\mu_c^*$ of 0.16. (These quantities are not available for the dynamically unstable structures with imaginary modes, for example Y$_2$C$_3$ $<$ 20 GPa, $\sigma$ $<$ 0.10 Ry and distorted structure with mode 1.) Experimental T$_c$ for Y$_2$C$_3$ ranges from 6 K to 18 K depending on synthesis conditions and measurement techniques \cite{KGKS69,AAMZA04}.}
    \begin{tabular}{|c|c|c|c|}
    \hline
        Stabilization & $\lambda$ & $\omega_{log}$ (K) & $T_c$ (K)\\
       \hline
         Mode 4 & 0.92 & 219.6 & 9.3\\
       Mode 10 & 0.93 & 219.4 & 9.5 \\
         
      20 GPa & 3.14 & 96.1 & 16.0 \\
         30 GPa & 1.38 & 263.6 & 22.3 \\
         
         $\sigma = $ 0.10 Ry & 1.14 & 228.5 & 14.5\\
         \hline
    
    \end{tabular}
   
    \label{tab1}
\end{table}

Besides distorting the $I$-$43d$ structure for more stable low-symmetry structures, increasing pressure and electronic smearing are other ways to shift the flat band away from E$_F$ (see Fig.A3 and A4) to stabilize the imaginary phonon modes. As plotted in Fig.~\ref{fig:4}(d), a large electronic smearing of 0.1 Ry on the $I$-$43d$ structure reasonably mimics the results of the low-symmetry $P1$ structure,as seen in phonon dispersion of Fig.~\ref{fig:4}(e) and spectral function ($\alpha^2$F($\omega$)) of Fig.~\ref{fig:4}(f). But pressure stabilizes these imaginary phonon modes with additionally interesting behaviors. As shown in Fig.~\ref{fig:4}(a) and Fig.A3(c), pressure of 20 GPa stabilizes the imaginary phonon modes, one at $\sim$4 THz and the other at as low as $\sim$2 THz, with unusually large $\lambda$ coming from those modes with large contribution from C dimers. The increasing pressure also shifts the flat band and the associated local DOS maximum peak first to cross E$_F$ and then to higher energy away from E$_F$, resulting in a smooth part of DOS at E$_F$, thereby stabilizing the imaginary phonon modes (Fig.A3(a)). As listed in Table.~\ref{tab1}, the large $\lambda$ $\sim$ 3.14 at 20 GPa and low $\omega_{log}$ $\sim$ 96.1 K are the results of the extremely soft or low-energy modes with a large EPC contribution, also seen in Fig.~\ref{fig:4}(c). For a further increase of pressure to 30 GPa, the lowest soft mode is pushed upward to bring the two stabilized phonon modes close to each other at $\Gamma$, similar to the cases of the $P1$ structure and the large electronic smearing, but achieving both a larger $\lambda$ of 1.38 and a larger $\omega_{log}$ of 263.6 K (Fig. A3(c)), thereby enhancing the T$_c$ from 9.3 and 14.5 K to 22.3 K. This trend of higher T$_c$ for smaller lattice constants under pressure is consistent with the experimental observation \cite{KGKS69,AAMZA04}, where an increase of T$_c$ from sub-10 K to 18 K has been observed when the lattice constant is decreased from 8.24 to 8.20\AA\ mostly due to different heat treatment and sintering conditions.

\section{Conclusions}
In conclusion, we find that the high-symmetry body-centered cubic (BCC) $I$-$43d$ structure of Y$_2$C$_3$ is dynamically unstable with zone-center imaginary optical phonon modes from C dimer wobbling motion due to the electronic instability from a flat band at the Fermi level (E$_{F}$). Once the imaginary modes are stabilized in a more stable low-symmetry structure, under pressure or with a large electronic smearing, these low-energy modes of mixed C and Y characters give arise to strong electron-phonon coupling (EPC) strength determining the superconducting properties. Our work demonstrates the necessity of further exploration in the compounds with imaginary phonon modes. EPC calculations may not always be feasible for the distorted structures due to low symmetry for large unit cells, however, pressure and electronic smearing are the alternatives to stabilize the imaginary phonon modes to study superconductivity. This can open a myriad of opportunities to discover high-temperature superconductors via high-throughput screening. Furthermore, we provide a thorough analysis of the phase stability of Y$_2$C$_3$. We have considered the different structural motifs of binary compounds containing group IV elements at the 2:3 composition and also explored energy landscapes via $ab$ $initio$ molecular dynamics near and out of the BCC structure. We plot their enthalpy vs pressure to confirm that the BCC-type structures with C dimers are indeed preferred in the low to medium pressure range. Because of the wobbling motion of the C dimers, there are many local energy minimums of the BCC-distorted structures with degenerated energies. Consequently, the ensemble average of numerous BCC-distorted structures with C dimer wobbling motion at finite temperature is expected to yield an overall BCC structure, consistent with what was observed in X-ray diffraction.

\section{Acknowledgements}
We acknowledges Dr. Lorenzo Paulatto for the helpful discussion of atom-projected phonon dispersion. This work was supported by Ames National Laboratory LDRD and U.S. Department of Energy, Office of Basic Energy Science, Division of Materials Sciences and Engineering. Ames National Laboratory is operated for the U.S. Department of Energy by Iowa State University under Contract No. DE-AC02-07CH11358.


%

\clearpage
\newpage

\section{Appendix}
Figure A1 illustrates the different views of the Y$_2$C$_3$ crystal. In the high-symmetry body-centered cubic (BCC) structure in space group $I$-$43d$ (220), each Y ion is enclosed by six C dimers arranged in a twisted prism configuration. Conversely, each C dimer is surrounded by eight Y atoms forming a cage comprised of two trapezoids. This arrangement results in a wobbling motion of the C dimer, contributing to its inherent instability. In Figure A2, phonon dispersion plots are presented for different structures of lattice distortion from different imaginary phonon modes, including Mode 10 (which exhibits the same total energy as Mode 4 mentioned in the main text) and Mode 1 (slightly higher in total energy). Figure A3 shows the influence of pressure on the electronic band structure and dynamic stability. Upon applying pressure exceeding 20 GPa, the flat band in close proximity to the Fermi level (E$_F$) undergoes an upward shift. This change causes the local maximum of the density of states (DOS) to shift towards the conduction band region, leaving behind a relatively smoother region at E$_F$. As the result, the electronic instability is gradually removed and the imaginary phonon modes become stabilized, which provides significant electron-phonon coupling (EPC). Finally, Figure A4 illustrates the shift of the flat band and local DOS peak away from the E$_F$ resulting from increased electronic smearing. After charge density is converged self-consistently with different smearing values, the DOS calculation is then performed with the tetrahedron method.

\begin{figure*}[h!]
\renewcommand{\thefigure}{A1}
    \centering
    \includegraphics[scale=0.7]{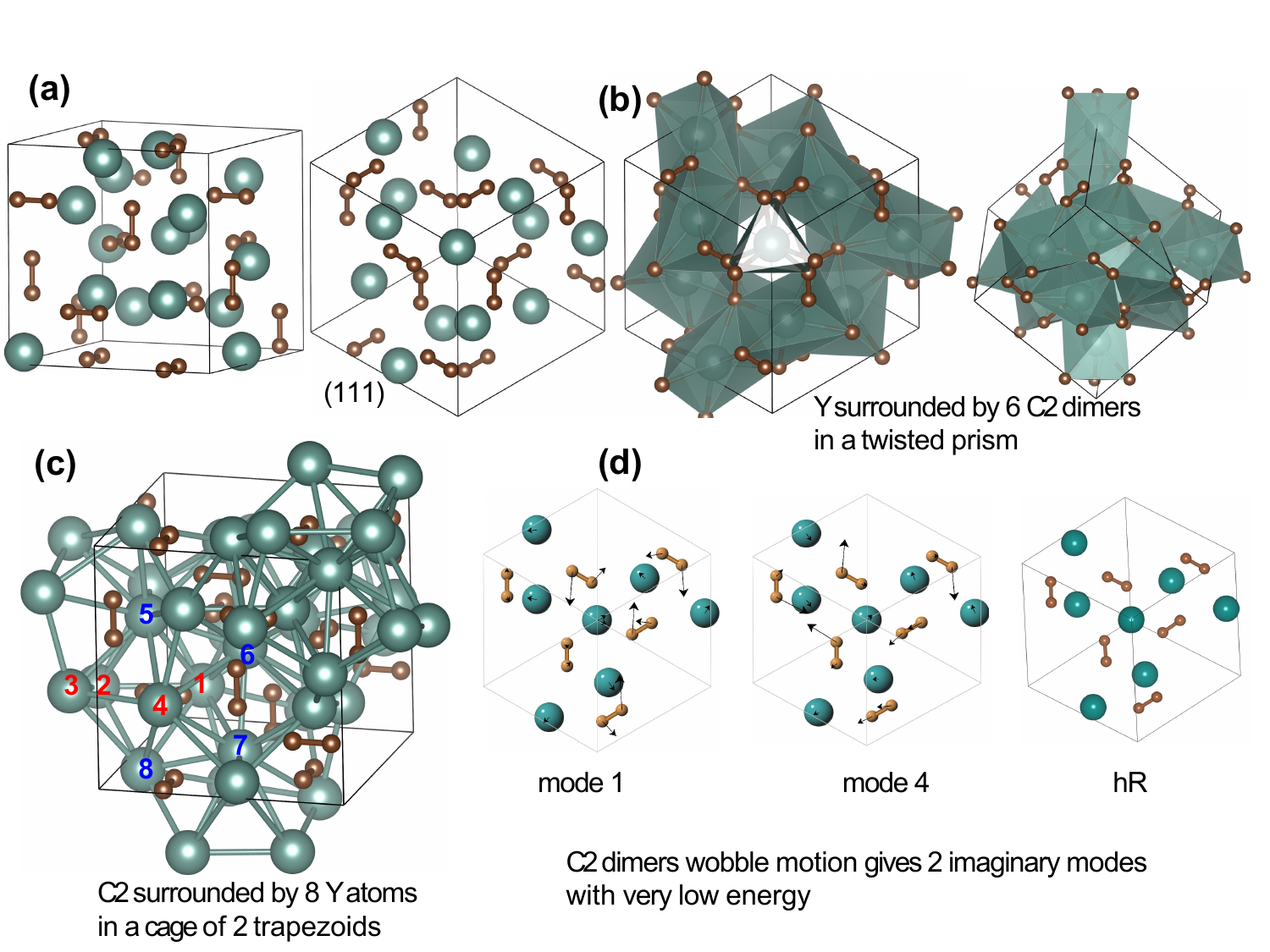}
    \caption{(a)-(c) Conventional body-centered cubic (BCC) unit cell of Y$_2$C$_3$ in space group $I$-$43d$ (220) viewed from different perspectives as labelled. Y and C atoms are represented by large and small spheres, respectively. (d) Atomic displacements from phonon eigenvectors plotted for mode 1 and mode 4; crystal structure of hR low-symmetry structure in space group $R3c$ (161).}
    \label{fig:S1}
\end{figure*}

\begin{figure*}[h!]
    \centering
    \renewcommand{\thefigure}{A2}
    \includegraphics[scale=0.7]{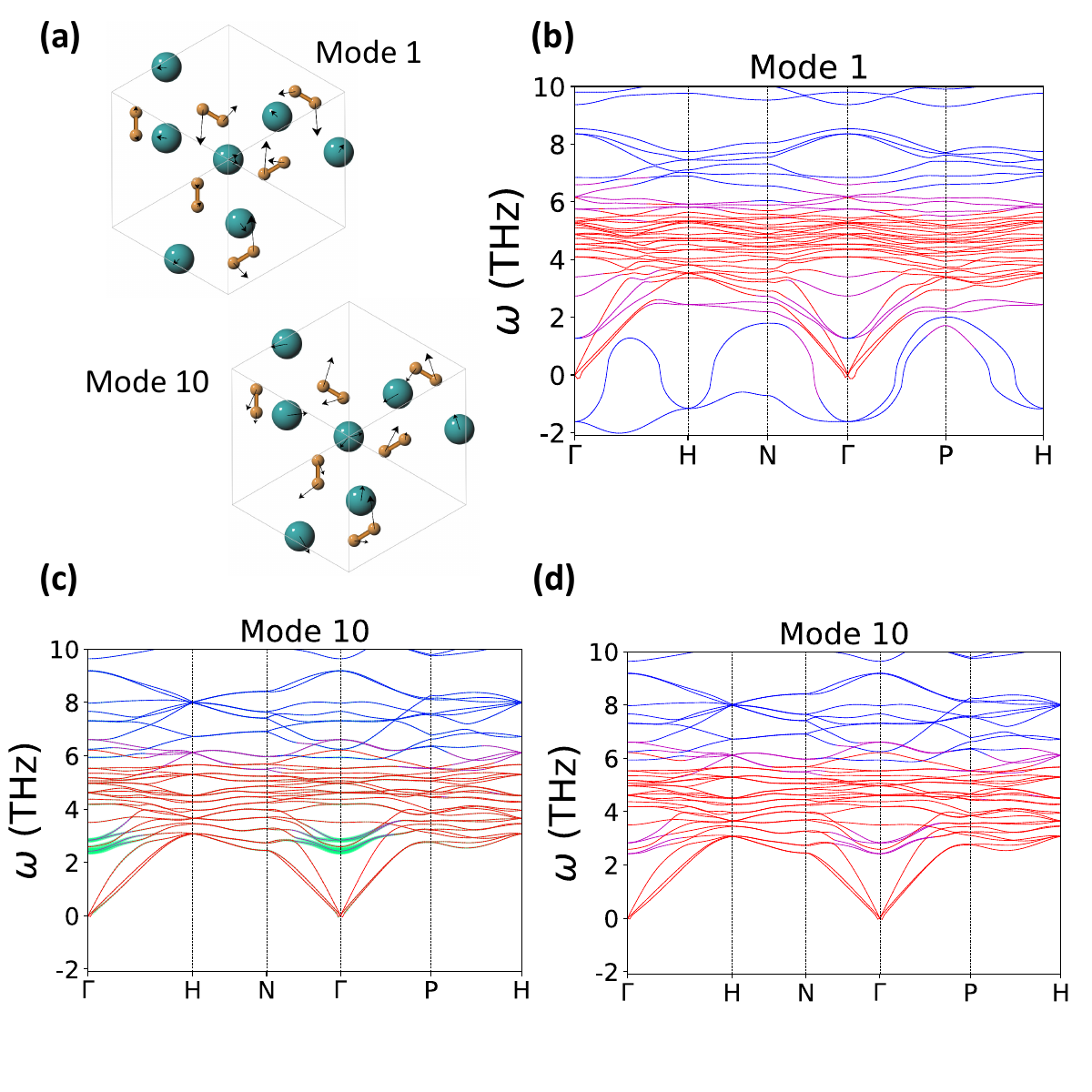}
    \caption{(a) Eigenmode plotted for modes 1 and 10 (b) Phonon dispersion showing imaginary phonon modes for the distorted lattice from mode 1. (c) Phonon dispersion for the distorted lattice from mode 10 (with same total energy as mode 4) indicating dynamic stability; electron-phonon coupling (EPC) strength is plotted (green color near $\Gamma$-point close to 2 THz phonon frequency). (d) Same plot as of (c) without EPC, but only with atomic character projection; purple color at the maximum EPC contribution indicates the contribution from both Y (red) and C dimer (blue); }
    \label{fig:S2}
\end{figure*}

\begin{figure*}[h!]
\renewcommand{\thefigure}{A3}
   \includegraphics[scale=0.45]{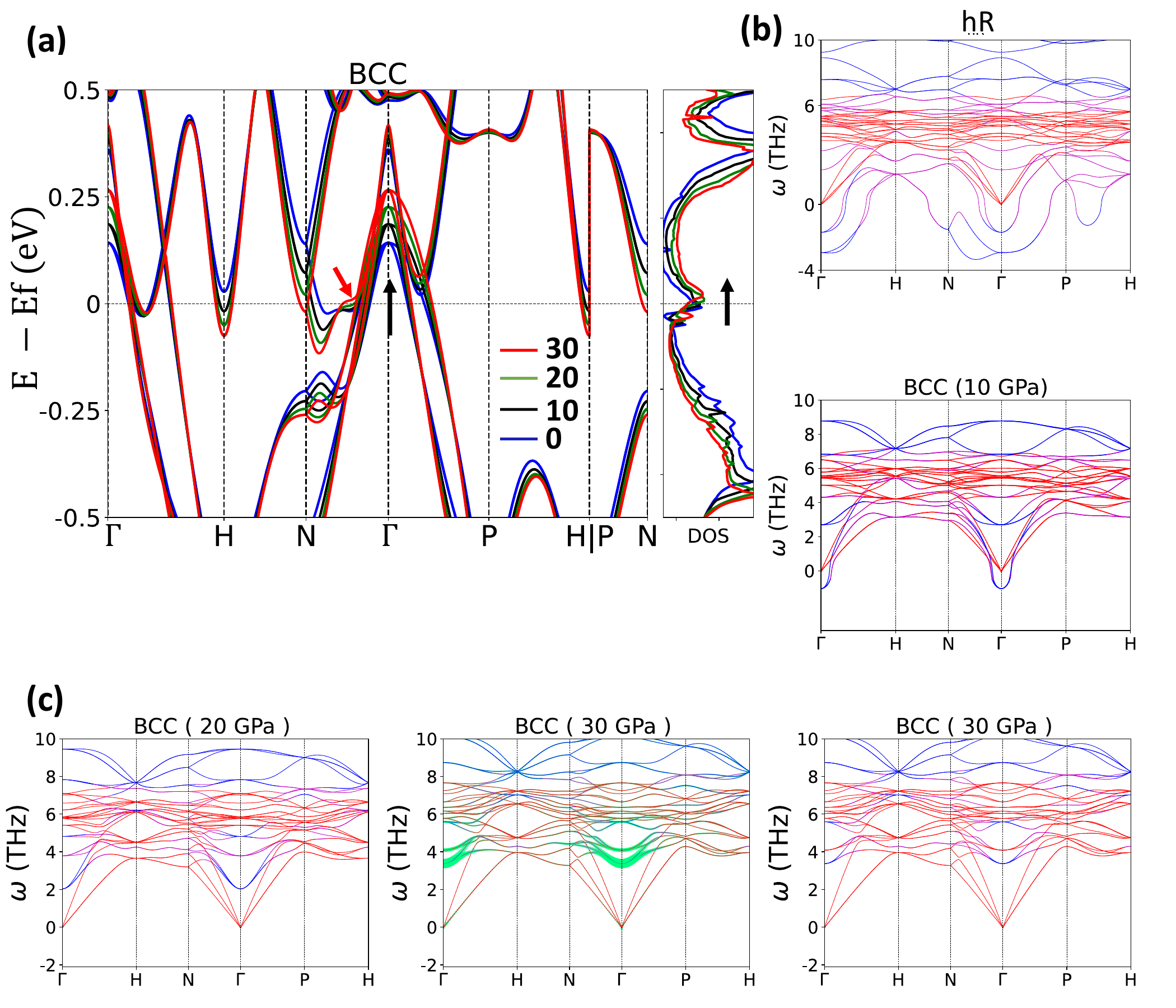}
  \caption{(a) Shifting of local density of states (DOS) maximum towards higher energy region with increasing pressure [from 0 GPa to 30 GPa], leaving relatively smooth DOS at E$_F$ (black arrow on right plot); A section of flat band near E$_F$ (shown by red arrow) is also pushed towards conduction band region (black arrow on left plot). (b) Phonon dispersion presenting dynamic instability of low-symmetry R3c structure and $I$-$43d$ (BCC) under 10 GPa pressure. (c) Phonon dispersion plots with projected atomic character for various pressure with/without EPC; two soft modes stabilized at 20 GPa are pushed upward with 30 GPa pressure.}
  \label{fig:S3}
\end{figure*}

\begin{figure*}[h!]
\renewcommand{\thefigure}{A4}
   \includegraphics[scale=0.45]{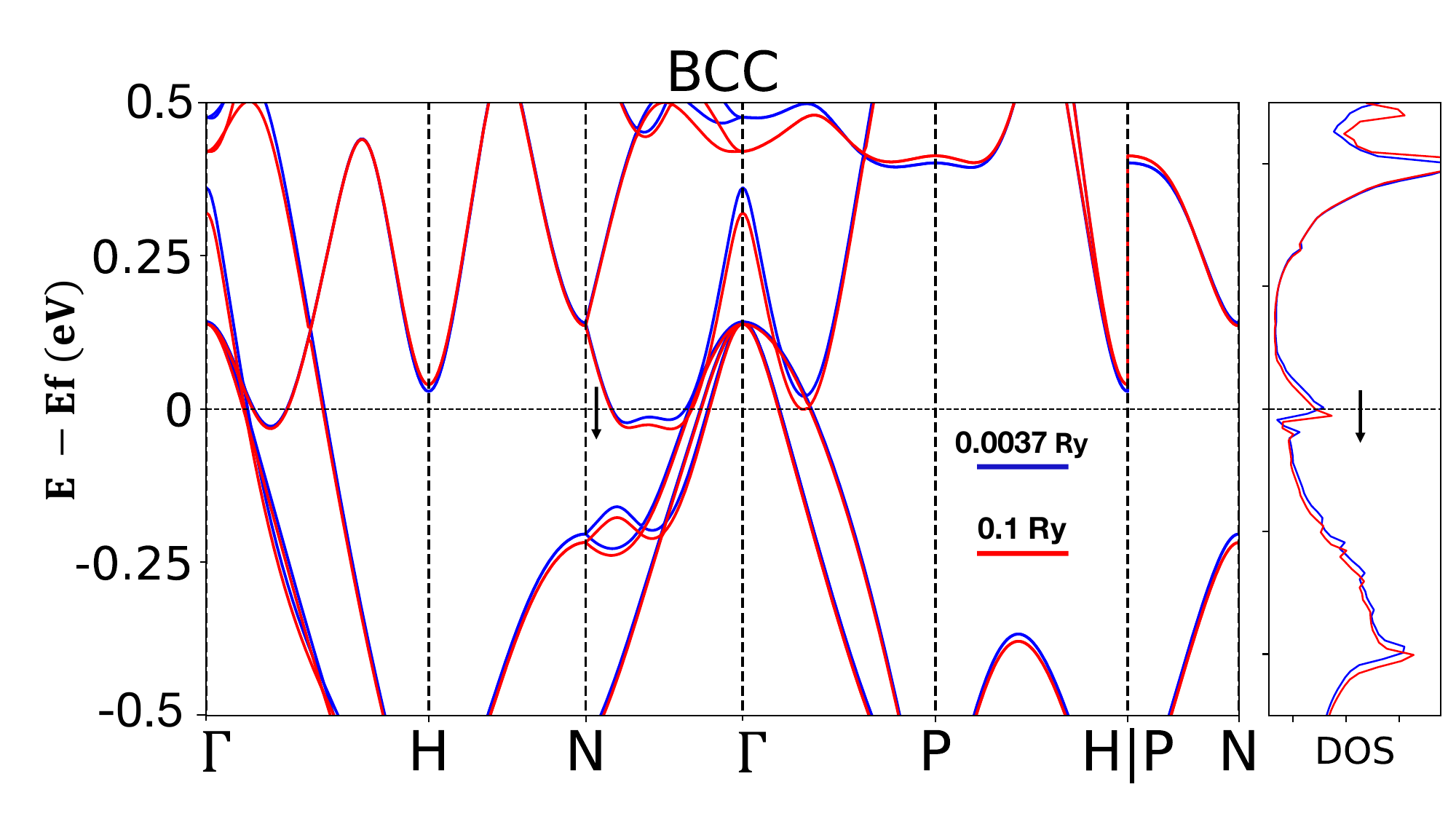}
  \caption{Shifting of flat band along the $\Gamma$-N and local density of states (DOS) maximum away from the E$_F$ towards lower energy with increased Gaussian smearing from 0.0037 Ry to 0.1 Ry. Note the DOS calculation is performed with the tetrahedron method, after the charge density is converged self-consistently with different smearing values.}
  \label{fig:S3}
\end{figure*}

\end{document}